# Bond disproportionation, charge self-regulation and ligand holes in *s-p* and in *d* electron ABX$_3$ perovskites by density functional theory

G. M. Dalpian,[1,2] Q. Liu,[1§] J. Varignon,[3] M. Bibes[3]
and Alex Zunger[1]

[1]University of Colorado Boulder Colorado 80309, Boulder, CO, USA
[2]Centro de Ciências Naturais e Humanas, Universidade Federal do ABC, 09210-580, Santo André, SP, Brazil
[3]Unité Mixte de Physique, CNRS, Thales, Université Paris Sud, Université Paris-Saclay, 91767, France

**Abstract**

Some ABX$_3$ perovskites exhibit different local environments (DLE) for the same B atoms in the lattice, an effect referred to as disproportionation, distinguishing such compounds from common perovskites that have single local environments (SLE). The basic phenomenology associated with such disproportionation involves the absence of B-atom charge ordering, the creation of different B-X bond length ('bond alternation') for different local environments, the appearance of metal (in SLE) to insulator (in DLE) transition, and the formation of ligand holes. We point out that this phenomenology is common to a broad range of chemical bonding patterns in ABX$_3$ compounds, either with *s-p* electron B-metal cations (BaBiO$_3$, CsTlF$_3$), or noble metal cation (CsAuCl$_3$), as well as *d*-electron cations (SmNiO$_3$, CaFeO$_3$). We show that underlying much of this phenomenology is the 'self-regulating response', whereby in strongly bonded metal-ligand systems with high lying ligand orbitals, the system protects itself from creating highly charged cations by transferring ligand electrons to the metal, thus preserving a nearly constant metal charge in different local environments, while creating B-ligand bond alternation and ligand-like conduction band ('ligand hole' states). We are asking what are the minimal theory ingredients needed to explain the main features of this SLE-to-DLE phenomenology, such as its energetic driving force, bond length changes, possible modifications in charge density and density of state changes. Using as a guide the lowering of the total energy in DLE relative to SLE, we show that

---

density functional calculations describe this phenomenology across the whole chemical bonding range without resort to special strong correlation effects, beyond what DFT naturally contains. In particular, lower total energy configurations (DLE) naturally develop bond alternation, gaping of the metallic SLE state, and absence of charge ordering with ligand hole formation.

## I. Introduction: Single *vs* multiple local bonding motifs for the same element in a crystal

Single repeated structural motif — be that $AX_4$ tetrahedron, or $AX_6$ octahedron, or $A_3B_3X$ trigonal prism – have established the basis of our understanding of structure and bonding in solids and molecules.[1–3] Furthermore, the tradition of using in electronic structure calculations the economically smallest possible unit cell, naturally forced in simple models the situation where each bonded element was described via a single local environment (SLE) -- the so-called *Monomorphous* representation. Ionic solids were generally modeled by the NaCl structure; intermetallic compounds by the $L1_0$ CuAu–type structure; and ternary $ABO_3$ oxides via the cubic perovskite ($Pm\bar{3}m$) structure. This view also underlies the description of disordered $A_xB_{1-x}$ alloys via the popular single site coherent potential approximation approach (CPA)[4,5] where all A atoms (and separately all B atoms) are assumed to see the same potential.

At the same time, the existence of more than one inequivalent Wyckoff position for identical elements in a lattice is no foreigner to crystallography. Classic example of polymorphous structures manifesting different local environments for the same chemical element involves elements capable of existing in multiple valences. For example, column III elements with the configuration $s^2p^1$ have, at low atomic number (Z), and for the top of the periodic table column (B, Al, Ga) a formal oxidation state (FOS) of 3, whereas for high atomic numbers, at the bottom of the column, the FOS might be 1 (e.g., Tl). The reason is that the relativistic Mass-Darwin effect[6] is sufficiently large to localize the $s^2$ electrons and make them quasi core–like orbitals, leaving a single $p^1$ electron at the high Z bottom of the periodic table column as chemically active. The intermediate Z elements — In and in part Tl — have two stable valences. An analogous transition occurs in column IV elements, where the light elements (Si, Ge) utilize all their four ($s^2p^2$) valence electrons, whereas the high Z elements (Pb) are mostly divalent ($p^2$) with the intermediate one (Sn) having two stable valences. In such dual valence atoms a single valence would disproportionate into two different valences, as illustrated by the "negative U center" of In in PbTe,[7] and by the dual valence of Sn in $Cu_2ZnSnS_4$.[8]

A particularly interesting case of different local environments to the same element is the disproportionation of a pair of identical $2\langle BX \rangle$ *single local environments (SLE)* in a $ABX_3$ perovskite into a structure with two *different local environments* (DLE) associated with the same element B, i.e., $\langle BX \rangle^{(1)} + \langle BX \rangle^{(2)}$ observed in $A_2[B^{(1)}B^{(2)}]X_6$ perovskites (such as $BaBiO_3$,[9,10] $YNiO_3$,[11] $CaFeO_3$[12,13]). Whereas the appearance of two different local environments is ubiquitous in *double perovskites,* when the disproportionating sites are *different* chemical elements as in $A_2[BC]X_6$ (for example $Sr_2[FeMo]O_6$[14] and $Cs_2[AgIn]Cl_6$[15]), in the current paper we discuss the unusual case of disproportionation with the same chemical element B.

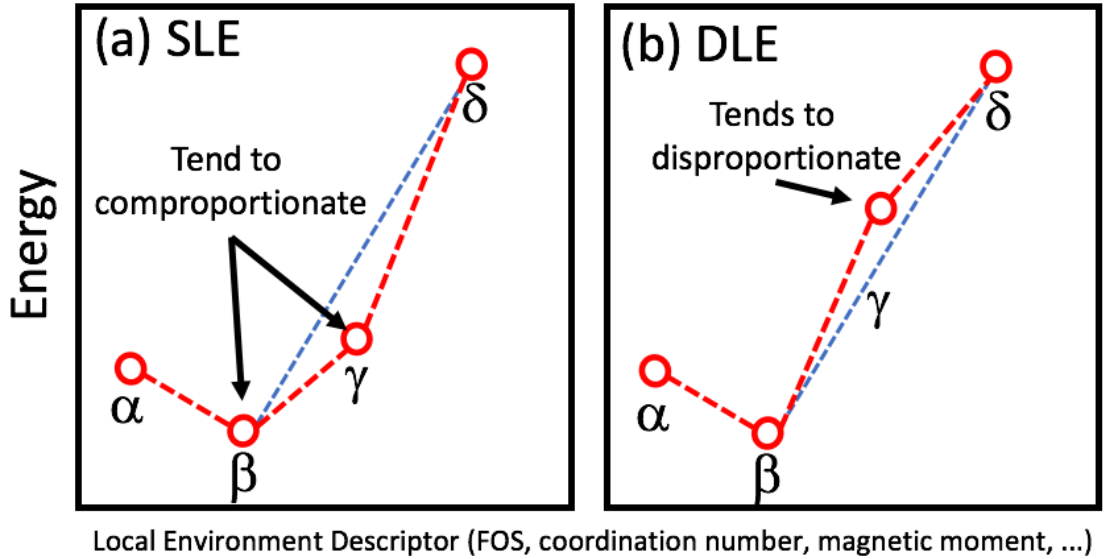

**Fig 1:** Schematic representation of a total energy of a fixed-composition compound $ABX_3$ appearing in a few hypothetical phases ($\alpha, \beta, \gamma, \delta$) of different local environments vs. the "local Environment Descriptor" which can, for example, be the formal oxidation state (FOS), or the coordination number, or the magnetic moment of the electronically active element. A phase below (a) or above (b) the tie line connecting the nearest neighbor phases (dashed blue line) will manifest in SLE and DLE behavior, respectively.

The existence of stable SLE *or* a stable DLE can be represented in a generalized 'Convex Hull' plot of total (free) energy of different phases *vs* some descriptor of the local environment (such as FOS, or coordination number, etc), as shown in Figure 1. In the first case (Fig. 1a), atom B in either the β or the γ state will not disproportionate because this would raise its energy relative to the 'tie line' represented by the straight blue line and will stay as an SLE, whereas in the second case (Fig. 1b) atom B in the γ state will disproportionate into β + δ because this lowers its energy relative to the tie line. We will follow in the present work this total energy guide for the tendency of various systems to manifest SLE or DLE behavior.

## II. The main questions addressed with respect to SLE *vs* DLE in ABX$_3$

We phrase below a number of questions posed regarding disproportionation in ABX$_3$ compounds, and will address them in this paper by considering six compounds showing DLE behavior, including BaBiO$_3$,[9,10,16–20] CsTlF$_3$,[21,22] SmNiO$_3$,[11,23–26] CsAuCl$_3$,[27] CaFeO$_3$[12,13,28–30] and CsTe$_2$O$_6$[31]. Although the last compound is not a perovskite, we also studied it to show that the conclusions drawn here can also be expanded to other types of structures. This selection represents a rather broad range of perovskite compounds with the B atom being an s-p element (Bi, Tl, Te), or a transition metal (Ni, Fe) or a noble metal (Au). We will demonstrate common behavior for all such cases, highlighting the broad appeal of SLE *vs* DLE selectivity.

*(a) What level of electronic structure theory is sufficient to predict the energetic tendency (Fig. 1) of actual ABX$_3$ compounds to be SLE or DLE?*

Previous studies implied that disproportionation is a correlation effect that may require an explicitly dynamically correlated approach. For example, Park *et al*[32] presented density functional plus dynamical mean field theory (DFT- DMFT) calculations "which show that the bond-length disproportionation and associated insulating behavior are signatures of a novel correlation effect".

Cammarata *et al*[33] proposed a 'spin assisted covalent bond formation' as a mechanism for DLE. However, this model cannot be general, since compounds like BaBiO$_3$, CsTlF$_3$, CsAuCl$_3$, and CsTe$_2$O$_6$, are not spin-polarized yet they have a DLE phase. Also, the spin-assisted mechanism cannot provide an understanding of the metal-insulator phase transition in CaFeO$_3$ and SmNiO$_3$ since (i) they transit from a paramagnetic metal to a paramagnetic insulator not showing net spin polarization and (ii) the Néel temperature is well below the MIT. The magnetic interactions cannot therefore be accounted for the MIT for these two compounds. In our case, DFT was able to predict both phenomena: we always observe a decrease in energy when going from SLE to DLE, and we also concomitantly observe the opening of a band gap for both spin-polarized and non spin polarized configurations. We note another DFT explanation by Mercy *et al*[26] who have recently provided a compelling evidence that the octahedra rotations are triggering the MIT in nickelates and $CaFeO_3$,[29] reproducing the experimental observations.

In the present study we find that SLE-DLE selectivity exists in *s-p* as well as 3*d* electron compound alike, and that density functional theory (DFT) suffices to correctly describe the energetic selection between SLE and DLE in all such compounds. This establishes such single determinant, mean field band theory as adequate tool for describing the phenomenology related to bond disproportionation, including magnetism[25] and defects in disproportionated structures.

*(b) Is the FOS a physically meaningful 'local environment descriptor' for predicting within the convex hull construct of Fig. 1 the tendency of actual $ABX_3$ to be SLE or DLE?*

The most basic understanding of the disproportionation problem suggests that when DLE occurs, the elements located at the B site in $ABX_3$ perovskites will have different FOS. In this view the meaningful descriptor of the local environment of the disproportionating B atom is the FOS. For $BaBiO_3$, for instance, the Bi site was said[10] to disproportionate to represent the two valences of the Bi atom, resulting in $Ba_2[Bi^{3+}Bi^{5+}]O_6$. The same view of charge ordering is often used for the other compounds, such as $Cs_2[Au^{+1}Au^{+3}]Cl_6$, $Ca_2[Fe^{3+}Fe^{5+}]O_6$, and $Cs_2[Te^{4+}Te_3^{6+}]O_{12}$.

In 1951, Frost[34] proposed a way to determine if a specific FOS of a certain element is stable or not by plotting the free energy versus the FOS[35,36] (a specific choice of a 'Local Environment Descriptor' in Fig.1), and learning what would be expected for this specific compound. These graphs were constructed by using solutions and electrode potential free energies. This view that integer oxidation states are physically realizable (as opposed to being formal labels) led to the picture equating such disproportionation to "charge ordering"[19,37,38] whereby the formal oxidation states corresponds to physical charges, alternating on the different chemically identical elements throughout the lattice.

To establish whether a structural change such as SLE-to-DLE is associated with a change in charge distribution we compute the quantity most directly related to our question, namely the variational charge density ρ(r) calculated self consistently by DFT for DLE and for SLE geometries. We therefore calculated the charge accumulation function i.e. the charge enclosed in a sphere of radius R around the B atom, as a function of R. Contrary to other methods of estimating the charge around an atom, such as Bader analysis, where a fixed boundary is chosen, the charge accumulation function is a direct measure of the charge density, providing direct evidence of the charge distribution around a certain atom. From these plots, it is also straightforward to clearly see the charge density difference around a certain atom. We clearly see that the physical charge density is essentially unchanged around the B atom as a result of the structural change. If one considers instead an indirect measure such as formal oxidation states, one deduces that it changes very significantly by the SLE- to- DLE transformation. We conclude that the FOS has little or nothing to do with the physical charge density. The reason for this was discussed in detail in Ref 39 in terms of the 'charge self regulating response', whereby charge rearrangement on the cation is offset by opposite rearrangement on the ligands, resulting in a minimal net change in physical charge density. In contrast, the FOS concept focuses just on the atom whose charge is counted, (namely, the B cation in the present case), seeing therefore just a piece of the picture. Similar conclusion were reached for the case of transition metal impurities in semiconductors,[39] $LiCoO_2$ vs $CoO_2$,[40] and for Sn atoms in perovskites such as $CsSnI_3$ and $Cs_2SnI_6$.[41]

*(c) Is the bond geometry a physically meaningful descriptor for predicting within the convex hull construct of Fig. 1 the tendency of actual $ABX_3$ to be SLE or DLE?*
Once it is understood that the charge residing in a certain B atom is basically constant for different local environments, it is important to look for a more relevant descriptor for this disproportionation. X-ray techniques can precisely determine the difference in bond lengths between the B atoms and the ligands. They can clearly differentiate the large and small octahedra in perovskites such as those studied in this paper.[10] While it is possible to assign different bond distances to different FOS,[9] the fact that the physical charge residing on different B atoms is nearly identical suggests that this assignment does not reflect a causal mechanism. For example, $PbCoO_3$ is said[42] to have both A-site and B-site charge ordering, leading to a formal description as $[Pb^{2+}Pb_3^{4+}][Co_2^{2+}Co_2^{3+}]O_{12}$. The characterization as a charge ordering compound in this case came from X-ray measurements that show two groups of Pb-O bonds and two groups of Co-O bonds and not the direct measurement of any quantity related to charge. Our analysis indicates that it is best to use the bond geometries to differentiate these DLE, since this is the property that is usually measured, and not the FOS. We will show that DFT can predict the observed bond disproportionation in all DLE compounds studied. Sawatzky and collaborators reached a similar conclusion for $BaBiO_3$[17] and Varignon *et al* for rare earth nickelates.[25] We therefore use the term 'bond disproportionation' rather than 'charge ordering/disproportionation'.

*(d) How is the SLE vs DLE selection related to metallic vs insulating character of the compound?*

It has been often observed that structural disproportionation comes with a simultaneous metal-insulator transition, e.g. in rare earth nickelates[11,26,32] and $CaFeO_3$[29]. For transition metal compounds, correlation effects have been used to explain the metal insulator transition.[32] We find in standard DFT description for both *s-p* $ABX_3$ systems and $ABX_3$ *d*-electron systems that whenever the SLE phase is metallic, the formation of the DLE configuration lowers the total energy (Viz. Fig 1b) and becomes automatically insulating. Specifically, in $RNiO_3$, the metal-insulator transition is developed by lattice mode couplings between rotations in DLE rather than by pure correlation.[26] Thus the metal-insulator transition is an energetic consequence of disproportionation in these systems.

*(e) How is disproportionation related to Ligand Hole?*

The basic electronic structure[43,44] of metal oxides involves a valence band maximum (VBM) made either of oxygen *p* orbitals (in late transition metal oxides such as NiO) or from metal atom *d* orbitals (in early 3d oxide compounds such as $VO_2$[45]). The conduction band minimum (CBM) of such metal oxides is generally composed of transition metal *d* orbitals (in early transition atom oxides such as $YTiO_3$) or metal *s* orbitals (in late 3*d* oxides, CdO). A special case is when the CBM is made of ligand orbitals, called "*ligand hole*" states.[17,18,25,29,43] Ligand holes have been shown to exist in disproportionated systems[46] but there seems to be significant lack of clarity if they are intimately related to *d* electron systems and if they are specific to disproportionated systems.

We demonstrate by DFT calculations that the conduction band wavefunction of both *s-p* electron and *d*-electron disproportionated $ABX_3$ compounds discussed here represent ligand hole states. As to the mechanism of LH formation, we note that this requires that the relevant metal states be deeper than ligand orbitals so electrons can be transferred to the metal (as in late but not early transition metal oxides, or Bi compounds with low s electron valence states) and that a sufficiently strong metal-to-ligand coupling exist so as to split the ligand VB into occupied and unoccupied parts, the latter being LH. Thus, LH does not require disproportionation, (indeed we find it to exist in SLE configurations) but in disproportionated states there exist a short enough B-Ligand bond that could facilitate splitting of the valence band and LH formation. The basic driving force for LH formation is the *self-regulating response*:[39] Total energy lowering favor the formation of LH when without such a LH, the charge on the metal would be very large (such as $Ni^{4+}$ in $RNiO_3$ or $Bi^{5+}$ in $BaBiO_3$), which is not favored energetically. Consequently, the ligand transfers electrons (thus, forming a hole) to the metal cation so as to self-regulate its charge (creating the $[Ni^{2+} - O^{1-}]$ complex in $RNiO_3$ and the $[Bi^{3+}-O^{1-}]$ complex in $BaBiO_3$, where the hole is localized on the oxygen octahedra). In addition, ligand holes were recently discussed in organometallic systems when strong π–π interaction splits the ligand π band so the upper π· band is unoccupied.[47,48]

We next provide a detailed discussion of the five questions above, leading to the conclusion that the phenomenology of disproportionation—absence of charge ordering, formation of bond length disproportionation, gap formation, as well as ligand hole formation— is derived and detected by total energy lowering within standard DFT, and is common to both s-p electron and d electron $ABX_3$ compounds.

## III. Method

***Density functional description of the SLE to DLE transformation:*** With the advent of accurate first principles exchange and correlation functionals[49] and effective energy minimization strategies (local gradients,[50] minima hopping,[51] Global Space Group Optimization,[52] etc.), the possibility of affording larger-than-minimal super cells, which provide an opportunity for chemically identical atoms to develop their own unique local environments has arisen. Consequently, it became possible to simulate this kind of SLE/DLE materials from a computational perspective, getting insights into the origin of these different configurations for the same atom.

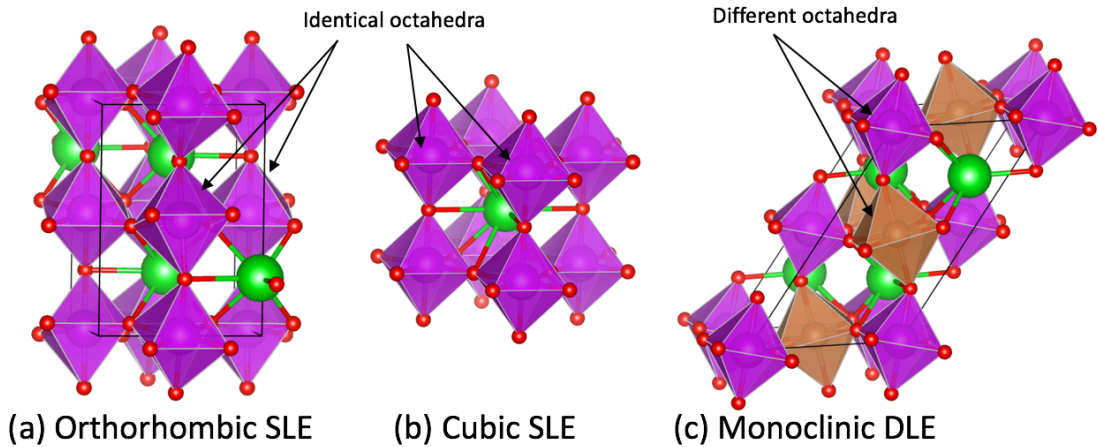

(a) Orthorhombic SLE     (b) Cubic SLE     (c) Monoclinic DLE

**Figure 2:** Geometrical representations of $ABX_3$ crystal structures with SLE and DLE. (a) is for an orthorhombic structure with SLE; (b) is a cubic structure with SLE; (c) is a monoclinic structure with DLE. In this case, the different colors of the octahedra indicate different local environments.

*Electronic Hamiltonian and its solver:* Calculations were performed using the plane wave pseudopotential total energy DFT approach as implemented in the VASP[53] code within the Projected Augmented Wave (PAW) approach and the Generalized Gradient Approximation (GGA-PBESol[54] for $CaFeO_3$ and $SmNiO_3$ and GGA-PBE[55] for the other materials). We have also performed hybrid functional calculations[56] on GGA-converged structures, in order to obtain total energies and band gaps. An on-site self interaction correcting "DFT+U"[57] term was added to $d$ orbitals of Ni (U=2.0eV) and Fe (U=3.8 eV). These values of U were chosen based on extensive tests from previous studies.[25] Basis set cutoff energies were set to 600eV for $CaFeO_3$ and $SmNiO_3$ and 400eV for the other compounds. The Brillouin zones were samples with k-point meshes up to 8x8x6 for orthorhombic phases (20 atoms) and 6x6x6 for cubic ones (5 atoms).

*Input crystal structures for relaxation:* When available, we have used crystal structures reported in the ICSD[58] for our calculations, and optimized both lattice vectors and internal atomic coordinates to minimize total energies until the forces on each atom for each Cartesian coordinate are smaller than 0.001eV/A. Such a relaxation scheme allows the system to change the symmetry of the trial structure. The most common structures observed in our SLE and DLE configurations are shown in Figure 2. In the SLE case (Figs 2a and 2b), all octahedra have the same shape and size. In the DLE configuration (Fig 2c) there are two different octahedra, arranged in such a way that a large octahedra is surrounded by six small octahedra and vice versa. Table I reports the space group symmetries of our optimized structures for both SLE and DLE phases, together with the references for experimental papers reporting these structures. For CaFeO$_3$, both SLE and DLE configurations have been observed experimentally (so the convex hull illustrated in Fig 1 must be rather shallow), the DLE being the low temperature structure.[13] For this case we use as trial structure the experimental crystal structure, with a ferromagnetic configuration for the spin arrangement, then relax the structure. For SmNiO$_3$ only the SLE configuration has been observed,[23] although our theoretical calculations show that the DLE configuration should be more stable. In this case, the initial DLE structure was copied from CaFeO$_3$, then fully relaxed. For other compounds (BaBiO$_3$, CsTlF$_3$ and CsAuCl$_3$ and CsTe$_2$O$_6$) only the DLE configuration has been experimentally identified so the trial SLE configuration was built by using either a cubic or an orthorhombic phase, inspired in other perovskites, followed by full relaxation.[59]

**Table I:** Calculated space group for the SLE and DLE compounds after DFT optimization. The experimentally observed configurations are marked in **bold**.

| Material | Space Group SLE | Space Group DLE | Reference |
|---|---|---|---|
| BaBiO$_3$ | Pnma | **P2$_1$/c** | 10 |
| CsTlF$_3$ | Pm$\bar{3}$m | **Fm$\bar{3}$m** | 21 |
| CsAuCl$_3$ | Pm$\bar{3}$m | **I$_4$/mmm** | 27 |
| CsTe$_2$O$_6$ | Pnma | **R$\bar{3}$m** | 31 |
| SmNiO$_3$ | **Pnma** | P2$_1$/c | 23 |
| CaFeO$_3$ | **Pnma** | **P2$_1$/c** | 13 |

### IV. Results

**A. Energy lowering due to disproportionation**

An interesting as well as pragmatic question is what is the minimum electronic structure theory framework needed to systematically predict spontaneous SLE/DLE symmetry breaking when it occurs? Although one can always go to the highest level methods to study certain materials, (such as quantum Monte Carlo,[60]) a reasonable question is what is the minimal set of physical ingredients that provide such a prediction, Recent publications claimed that the band gap opening for the DLE phase, for instance, is a strongly correlated effect,[32] and that methodologies such as Dynamic Mean Field Theory should be used to describe it. It turns out that, as Table II shows, the single determinant, mean field Bloch periodic DFT band theory with appropriately flexible unit cell that permits symmetry breaking if it lowers the energy, is essentially sufficient, at least for the (rather chemically broad) set of representative compounds used here.

**Table II:** Energy difference between DLE and SLE phases ($\Delta E_{(DLE-SLE)}$) for the studied materials. The band gaps have been calculated both within the GGA and HSE (in parenthesis) functionals for the DLE phase.

| Material | $\Delta E_{(DLE-SLE)}$ (meV/f.u.) | DLE Band Gaps (GGA(HSE) eV) |
|---|---|---|
| $BaBiO_3$ | -107 | 0.00 (0.63) |
| $CsTlF_3$ | -40 | 0.91 (1.78) |
| $CsAuCl_3$ | -732 | 0.70 (1.51) |
| $CsTe_2O_6$ | -136 | 0.38 (1.59) |
| $SmNiO_3$ | -82 | 0.05 (1.50) |
| $CaFeO_3$ | -4 | 0.09 (0.77) |

The energy lowering from SLE to DLE for all compounds in Table II indicate that the DLE phase is lower in energy than the SLE one. This is a clear indication that DFT can predict DLE configurations, and should be a good choice for studying this kind of phenomena. In the section D, we will show that DFT is also enough to describe band gap opening in these compounds, leading to a complete description of them.

**B. The physical charge density around the different disproportionated atoms is nearly constant thus no 'Charge Ordering'**

It was once thought[12,13,21,28,61] that disproportionation leads to different *physical charges* on the different B atoms in the lattice, an effect referred to as 'charge ordering'. This kind of phenomena has been labeled in several different ways such as charge ordering,[62] charge disproportionation,[13] valence-skipping,[20] missing valence states[63] or mixed valence compounds.[22,64] This view resulted from confusing formal charges assigned on the basis of the extreme ionic view, with physical charge observed in the variational charge density ρ(r). A few recent works have challenged the existence of charge ordering,[17,65] suggesting that bond disproportionation, or different local environments, should be a better description of the physical reality.

We have analyzed the variational DFT calculated charge density profiles around the B atoms for six structurally relaxed compounds. In Fig 3 we plot the total valence charge density of these compounds in a few different ways. First, in the left panels, we present a 2D representation of the total charge density in a plane containing both B atoms (inside small ($B_S$) and large ($B_L$) octahedra) of the DLE phase. The figures in the center panel show the charge density along a line containing both B atoms, as indicated in the left panels, reporting a very small difference in the absolute values for the charge residing in both. The "charge accumulation function" $Q(R) = \int_0^R \rho(r)dr$ integrated in a sphere of radius R around each B atom is shown in the figures on the right panels as a function of the sphere radius R. We note again a minimal difference between $Q(R)$ of SLE and DLE phases, and a larger difference is seen as R approaches the ligands, far from the B atom. *For all the studied cases we find that the charge around the B atoms in different local environments, supposedly designated by widely different oxidation states, is rather similar.*

This principle of conservation of cation charge under different bonding conditions in strongly coupled metal-ligand compounds has been discussed in the context of the Self Regulating Response[39–41,66] for the case of transition metal impurities in semiconductors,[39] for Co in $LiCoO_2$ vs delithiated $CoO_2$,[40] and for Sn atoms in perovskites such as $CsSnI_3$ and for its reduced form where 50 % of the Sn is removed as in $Cs_2SnI_6$. The description above is a clear confirmation that the use of 'valence' or 'charge' for differentiating both atoms in different local environments is not a good choice, since the charge in both is basically the same. Although there might exist a very small difference between the charge density on the two atoms, this difference is far from the two electrons argued by the valence skipping proposals. Such behaviors were explained earlier[25,39–41,66] by the cooperation of the ligand orbitals that rehybridize in response to a change in total charge (reduction; de lithiation; charging a sample) so as to minimize the perturbation—a manifestation of the Le Chatellier principle.

Given that the physical charge on the disproportionated atoms is rather similar, the next obvious question is: "What is a physically meaningful descriptor of the local environment of the disproportionating B atom?" This will explain, via Fig.1, which compound disproportionates and which stays as an SLE. The answer, as discussed next, is the bond geometry around each of the disproportionated atoms.

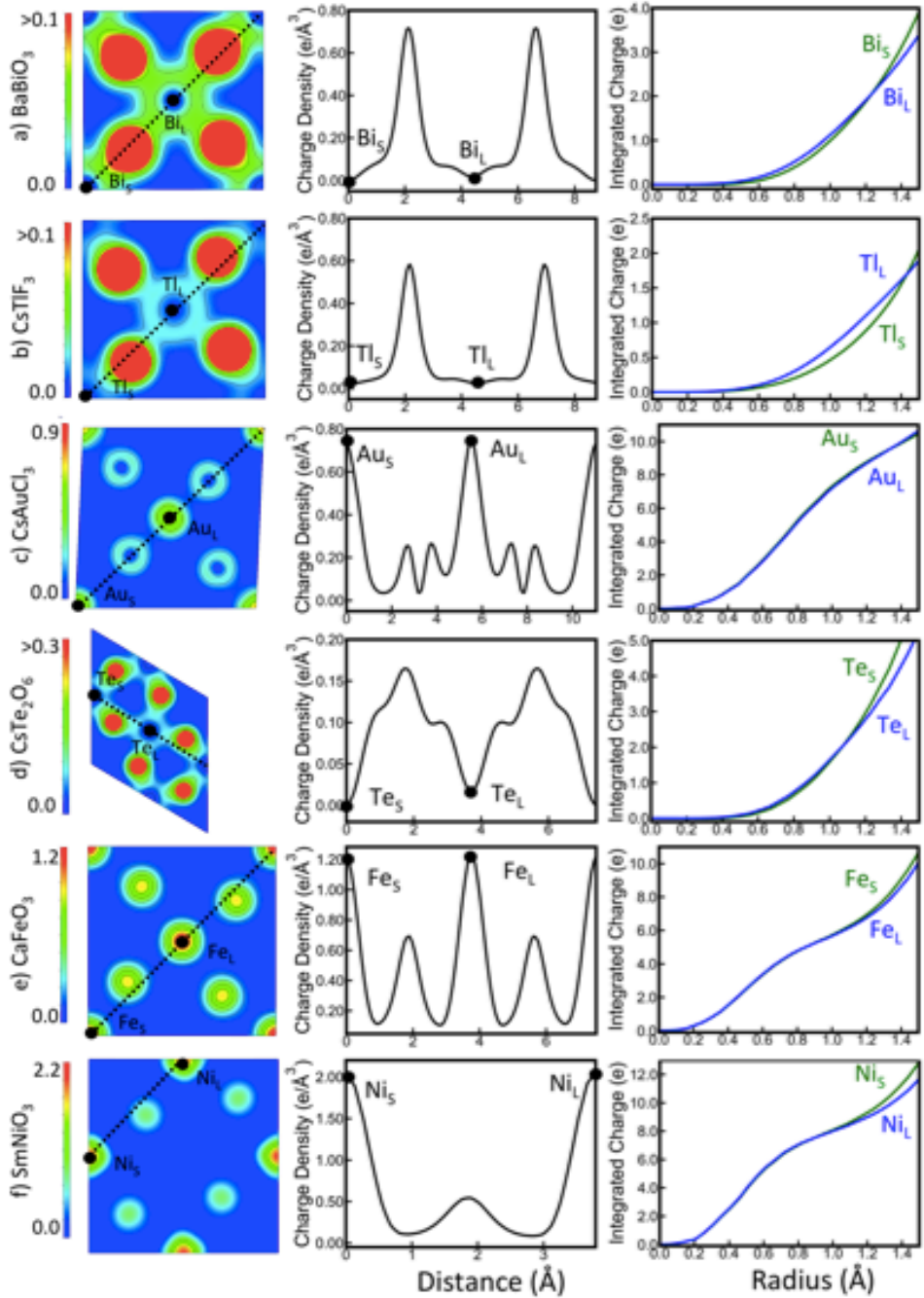

**Figure 3:** Charge density profiles for the DLE phase of (a) $BaBiO_3$, (b) $CsTlF_3$, (c) $CsAuCl_3$, (d) $CsTe_2O_6$, (e) $CaFeO_3$ and (f) $SmNiO_3$. The images on the left represent a 2D plot of the total charge density in a plane containing the B atoms in the small ($B_S$) and large ($B_L$) octahedra (in units of $e/Å^3$). The central figures show a 1D plot of the total charge density along the line shown in the figures in the left. The graphs in the right represent the total charge density integrated in a sphere of radius R centered in the $B_S$ and $B_L$ atoms, as a function of R (charge accumulation function). In all these figures, it is possible to observe that the charge around both B atoms ($B_S$ and $B_L$) is basically the same.

## C. The Different B-X bond lengths in DLE and SLE octahedra form good markers for disproportionation

A good way to differentiate the B atoms at different local environments in DLE compounds is through a directly measurable quantity such as the bond distance between the B and X atoms. Figure 4 shows the calculated B-X bond lengths in SLE and DLE phases of each of the studied materials, and a comparison with the available experimental results. For the SLE phase (red bars in Fig 4), the equal octahedra can either have six equal bonds, as in the case of $BaBiO_3$, $CsTlF_3$, $CsAuCl_3$ and $CsTe_2O_6$, or be distorted with two different bond lengths, such as in the case of $CaFeO_3$ and $SmNiO_3$. For the DLE compounds, the small octahedra are represented by green bars, and the large octahedra are represented by blue bars. For the DLE cases, owing to the lower symmetry of the compounds, there can be different groupings of bond lengths in each octahedra. Experimental bond distances are shown in Fig 4 as white circles.

Although there are claims stating that standard DFT "strongly underestimates the breathing distortion parameters" in DLE configurations,[16] it is clear from Fig 4 that DFT provides a good description (see % deviation listed in Fig 4) of the trends on bond lengths upon disproportionation across the different bonding groups, whether the active B cation is d-electron, or s-p electron or noble metals. An outlier is $CsTlF_3$ with a deviation of 6% which might be related to sample stability/quality concerns reported in Ref. 21 . While future improved DFT functionals could hopefully improve the quantitative agreement with experiment, there is little doubt that even current DFT functionals provide already a reasonable picture of SLE-to-DLE spontaneous symmetry breaking.

.

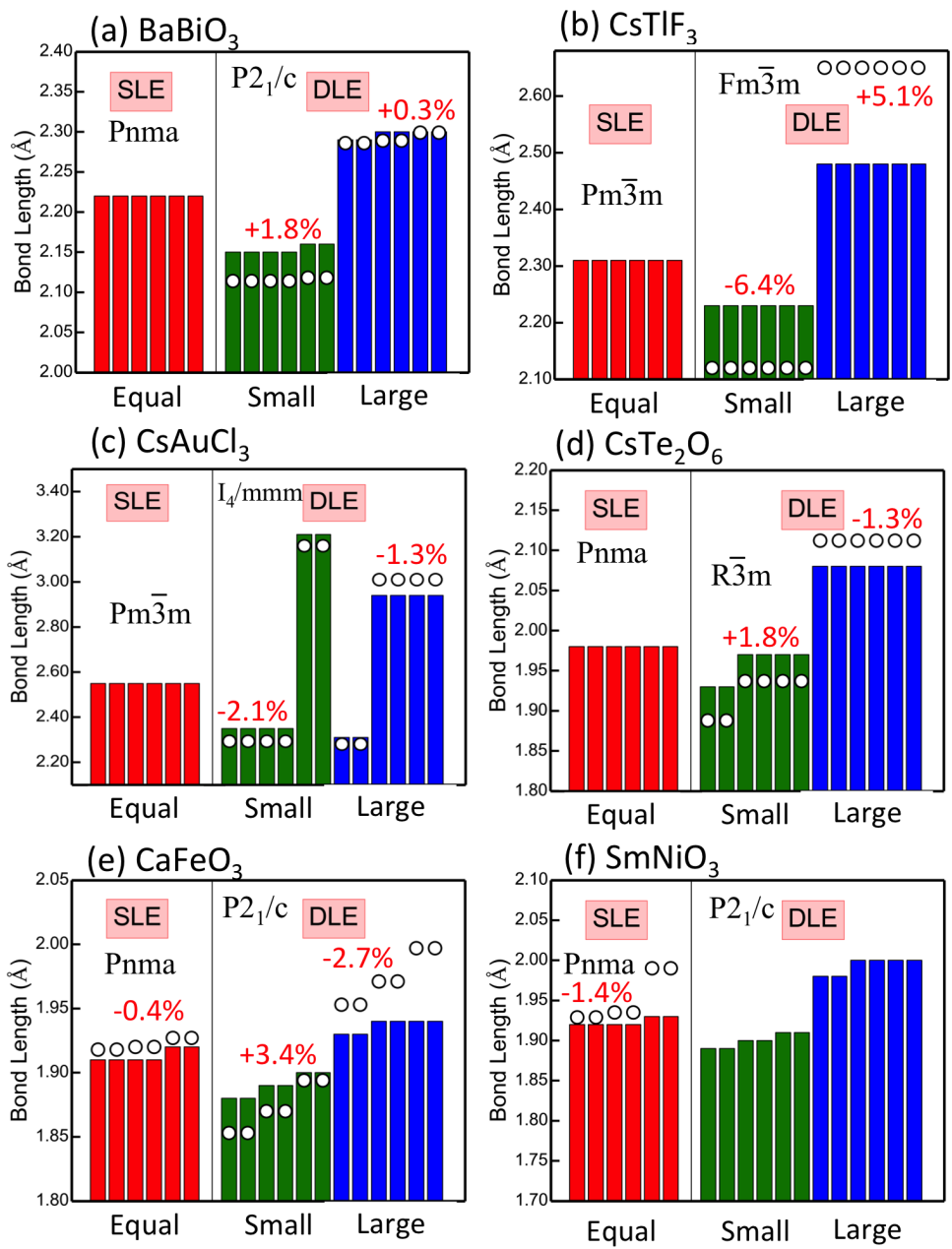

**Figure 4:** Bond lengths for the studied materials in both SLE and DLE configurations. For the DLE phase, distances are depicted for both the large (blue) and small (green) octahedra. For SLE, all octahedra are the same (red). White circles indicate experimental measurements. The calculated space groups and the average difference between experiment and theory are indicated. Note different scales on different graphs.

Besides the bond distance analysis shown in Fig 4, we have also performed a more detailed comparative analysis of the magnitude of the different symmetry-allowed normal modes[67] including octahedral tilting (OT), Octahedral rotation (OR), breathing modes (BM) producing the rocksalt like pattern of compressed and extended oxygen cage octahedral, and oxygen anti polar (O-AP) displacements $X_5^-$ and $R_4^-$ for our compounds. The results presented in Table III correspond to the actual amplitude of each lattice distortion appearing in the ground state with respect to a high symmetry cubic phase. In other words, atomic displacements are decomposed on the basis of the eigenvectors of the Dynamical matrix (phonon modes) of the perfectly symmetric structure of perovskites ($Pm\bar{3}m$) phase, therefore providing the amount of each pure lattice distortions appearing in the material. The total amount of atomic displacements is reported in Angstroms. This analysis shows that the DFT calculated structures are, in most cases, in very good agreement with experiment.

**Table III:** Amplitudes of key distortions (in Å) appearing in the ground state of each materials (both SLE and DLE) obtained on the basis of a symmetry adapted mode analysis with respect to a high symmetry cubic phase. The analysis is performed with AMPLIMODES from the Bilbao Crystallographic Server. Only Octahedra rotations (OR and OT), the Breathing mode (BM) and anti-polar (O-AP) displacements are reported.

|  | DLE | | | SLE | |
|---|---|---|---|---|---|
| $BaBiO_3$ | $P2_1/c$ | | | | |
|  | Expt. | Theory | | | |
| BM ($R_2^-$) | 0.308 | 0.245 | | | |
| OR ($R_5^-$) | 1.226 | 1.370 | | | |
| O-AP ($X_5^-$) | 0.178 | 0.229 | | | |
| OT ($M_2^+$) | 0.384 | 0.414 | | | |
|  | | | | | |
| $CsTlF_3$ | $Fm\bar{3}m$ | | $P2_1/c$ | | |
|  | Expt. | Theory | Theory | | |
| BM ($R_2^-$) | 0.648 | 0.373 | 0.441 | | |
| OR ($R_5^-$) | - | - | 1.484 | | |
| O-AP ($X_5^-$) | - | - | 0.615 | | |
| OT ($M_2^+$) | - | - | 0.938 | | |
|  | | | | | |
| CsAuCl3 | I4/mmm | | | | |
|  | Expt. | Theory | | | |
| BM ($R_2^-$) | 0.256 | 0.125 | | | |
| O-AP ($R_3^-$) | 0.949 | 0.839 | | | |

| CaFeO$_3$ | P2$_1$/c | | | | Pnma |
|---|---|---|---|---|---|
| | Expt. | Theory | | | Theory |
| BM (R$_2^-$) | 0.180 | 0.081 | | | - |
| OR (R$_5^-$) | 1.078 | 1.091 | | | 1.084 |
| O-AP (X$_5^-$) | 0.406 | 0.461 | | | 0.456 |
| OT (M$_2^+$) | 0.833 | 0.798 | | | 0.938 |
| O-AP (R$_4^-$) | 0.176 | 0.081 | | | 0.091 |
| | | | | | |
| SmNiO$_3$ | P2$_1$/c | | | Pnma | |
| | | Theory | | Expt. | Theory |
| BM (R$_2^-$) | | 0.166 | | - | |
| OR (R$_5^-$) | | 1.277 | | 1.244 | 1.175 |
| O-AP (X$_5^-$) | | 0.616 | | 0.525 | 0.526 |
| OT (M$_2^+$) | | 0.914 | | 0.983 | 0.802 |
| O-AP (R$_4^-$) | | 0.140 | | 0.188 | 0.116 |

**D. Energy lowering upon DLE formation is accompanied by gaping and metal to insulator transition**

In some compounds like CaFeO$_3$[29] and in some rare earth nickelates such as YNiO$_3$[26] the structural transformation from DLE (an insulator) to SLE (a metal) configuration with increasing temperature is accompanied by an insulator to metal transition. The fundamental origin of the band gap opening as well as for the transition from SLE to DLE are still a matter of debate.

Table II shows the calculated band gaps both within the GGA approximation (or GGA+U for transition metal compounds) and also using HSE hybrid functionals for the studied compounds. HSE calculations consistently give larger band gaps for the compounds, as expected. For BaBiO$_3$, as previously discussed in the literature, the GGA band gap is zero. As the VBM and CBM are in different points of the Brillouin zone,[68] and there are no levels crossing the Fermi energy, this zero gap should not be a major problem for the analysis reported below.

Figures 5 and 6 show the density of states of the selected compounds in both SLE and DLE configurations. Figure 5 reports results for compounds that are not spin polarized, whereas in figure 6 we report the density of states for the magnetic compounds (only spin-up), where the ferromagnetic configuration was assumed. Other complex magnetic configurations might exist in these compounds,[25] but they will not be discussed in the present paper. Our test calculations with AFM configurations show that our main conclusions will not change with a different magnetic configuration.

As can be observed on the left panels of figures 5 and 6, all the SLE configurations are metallic. The red curve represents the ligand (Oxygen, Fluorine or Chlorine) states, whereas the green curves represent the B atom orbitals. For the DLE configuration, right panels on Figs 5 and 6, we separate the contribution from the B atom inside the large octahedra (painted green) and from the B atom inside the small octahedra (painted purple). Red curves again represent the ligands. The levels related to the A atom do not appear in the selected energy range. We can clearly see a specific qualitative behavior in these compounds: the levels related to the B atom in the large octahedra are mostly localized in the valence band, whereas those related to the B atom on the small octahedra are in the conduction band. As the coupling between the B atom and the ligands is larger in the small octahedra, these levels are pushed to higher energies (purple curves), whereas those from the large octahedra are mostly filled in lower energies (green curves). The different coupling between large and small octahedra is a clear and straightforward explanation of why the DLE phase is insulating.

The hybridization between B atom and ligands also leads to a large DOS contribution from ligand (oxygen, fluorine and chlorine) atoms around the Fermi energy. As shown in Fig. 5 and 6, for $BaBiO_3$, $CaFeO_3$ and $SmNiO_3$, the ligand contribution to the CBM is much larger than that from the B atom, consistent with the previous observation/prediction[17,29,43], while for $CsTlF_3$ and $CsAuCl_3$, the CBM has almost a similar contribution from the ligand and the B atom. We will further discuss such ligand hole states in Sec. F.

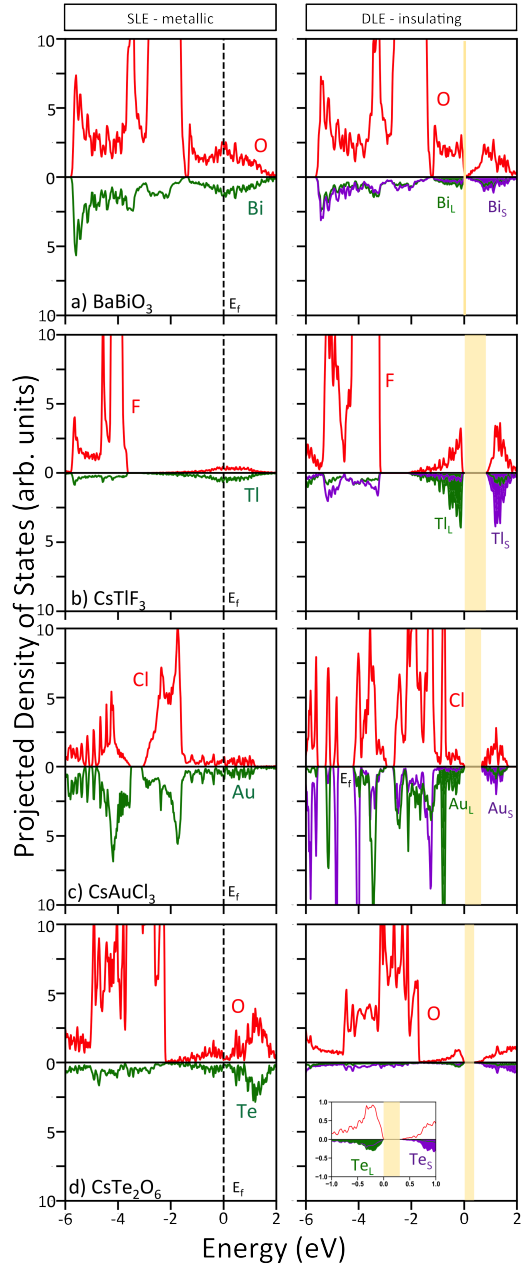

**Figure 5** Projected density of states for (a)BaBiO$_3$, (b) CsTlF$_3$, (c)CsAuCl$_3$ and (d) CsTe$_2$O$_6$. The figures in the left are for the metallic SLE configuration, where the green curve represents the B atom and red is related to the ligands (Oxygen or Fluorine or Chlorine). The figures in the right are for the insulating DLE configuration, where green refers to B atoms inside large octahedron, and purple to B atoms inside the small octahedron and the yellow region indicates the band gap. All these PDOS were calculated using GGA. The inset shows details of the region around the band gap.

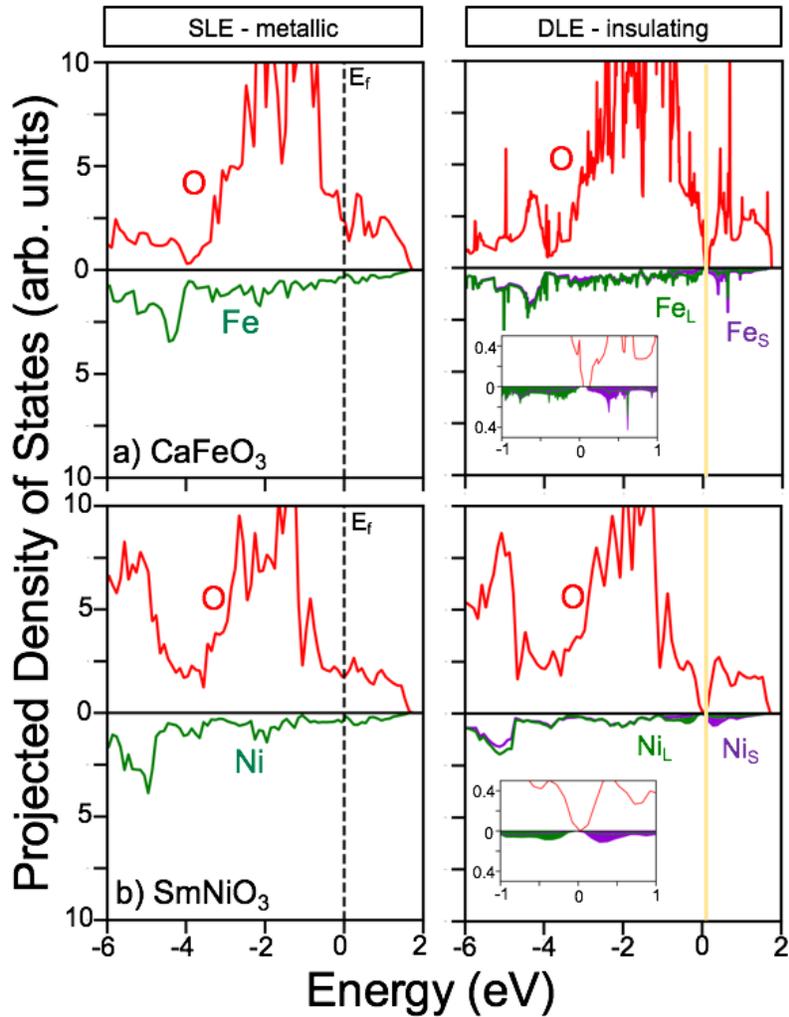

**Figure 6** Projected density of states for (a)CaFeO$_3$ and (b)SmNiO$_3$. The figures in the left are for the metallic SLE configuration, where the red curve represents the oxygen levels and green is for the B atom. Only spin-up components are plotted. The figures in the right are for the insulating DLE configuration, where green refers to the B atoms inside large octahedron, purple to the B atom inside the small octahedron and the yellow region indicates the band gap. The insets show details of the region around the band gap.

**E. Model for energy lowering and band gap opening in SLE- to-DLE conversion:**

The fact that for all studied compounds there is energy lowering and band gap opening when going from SLE to DLE is a clear indication that there is a universal behavior in these phenomena and, as such, it should be possible to develop a unified model to explain such properties. This can be done through an energy level diagram, as shown in Figure 7. In a first approximation, considering only the electronic contribution to the total energies, the energy lowering and band gap opening in DLE configurations can be understood through the different strengths of coupling between the B and X atoms in the $BX_6$ octahedra. In figure 7 we will use $BaBiO_3$ as an example, although similar trends can be extended to all other compounds.

*SLE bonding:* In Fig 7a, the coupling between oxygen $2p$ and bismuth $6s$ and $6p$ levels is depicted for the SLE configuration. As can be observed, this will lead to a metallic configuration, basically owing to electron counting. As all octahedra are similar, there will be only one strength for the coupling between the B atom and the X octahedra.

*DLE bonding:* In Fig 7b we show the same diagram for the DLE case. As there are now two different octahedra, one large and one small, we have to separately consider both of them. The coupling between Bi and O levels in the small octahedra is stronger, owing to the shorter bond distance. This will push the hybrid levels of the small octahedra upwards, emptying one s-p hybrid orbital. This empty hybrid orbital is usually called a *ligand hole* orbital, as will be discussed in the next section. For the large octahedra the coupling is weaker, leading to weaker repulsion. By considering both large and small octahedra, the final effect is that holes are pushed to higher energies and electrons are pushed to lower energies, leading in first order to an electronic energy gain. If the difference in the strength of the coupling for small and large octahedra is large enough, this will also open a band gap in this material.

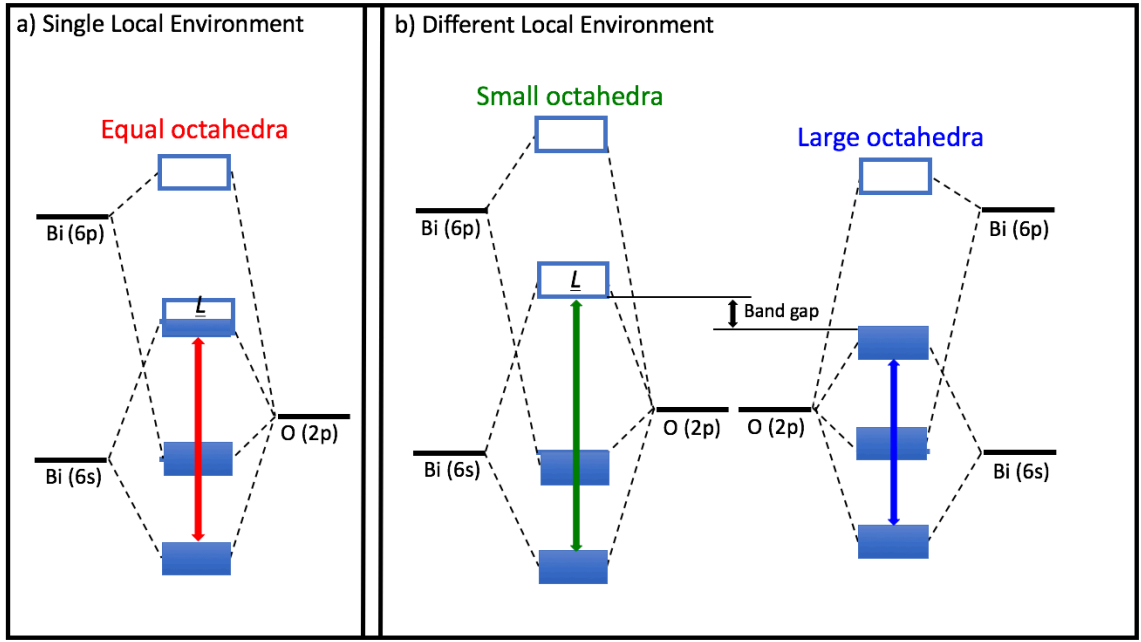

**Figure 7:** Schematic diagram showing the electronic coupling between B and X atoms in $ABX_3$. Here we use $BaBiO_3$ as example. (a) in the SLE configuration all octahedra have the same size. The coupling between Bi and O levels leads to a metallic configuration and the creation of a ligand hole ( $\underline{L}$ ). Filled (empty) boxes represent filled (empty) hybrid orbitals. (b) in the DLE configuration the octahedra have two different sizes leading to different strengths in the coupling between Bi and O. Small octahedra have a stronger coupling (green) and large octahedra have a weaker coupling (blue). This different coupling pushes the hybrid levels to higher energies, opening a band gap and lowering the total energy of the system.

The coupling diagram described in Fig 7 can also be transported to all other compounds, as can be clearly observed in their projected density of states in figures 5 and 6: the VBM of the DLE phase is always localized in the large octahedra, whereas the CBM is related to the small octahedra. For compounds containing d electrons, each material can have a different order of *e* and *t* levels, and the coupling is slightly different, although it leads to similar conclusions of energy lowering and band gap opening.

**F. Ligand holes in DLE compounds with strong metal-ligand bonds signal a self regulating response**

The conduction band of metal oxides (e.g. Fig 8a for NiO) is generally composed of either transition metal *d* orbitals (in early transition atom oxides such as $YTiO_3$) or from metal *s* orbitals (in late *3d* oxides). A special case is when the CBM is made of ligand orbitals, called "*ligand hole*" states.[17,24,29] Ligand holes have been shown to exist in disproportionated systems[46] but there seems to be significant lack of clarity if they are intimately related to *d* electron systems and if they are specific to disproportionated systems. The existence of ligand holes has been often associated with superconductivity in oxides,[69–71] making its understanding even more interesting.

***DFT Evidence for Ligand holes in s-p and d electron ABX$_3$ with strong metal-ligand bonding***: Fig 5 and 6 show the density of states of the empty conduction band indicating a clear ligand (oxygen) component. Figure 8 shows a 2D representation of the CBM charge density in a plane containing four X ligands and the B atom in the small octahedra for the studied compounds. For guiding the eye, we first show in Fig 8a the conduction band wavefunction square of the NiO system that *lacks* ligand holes. It is very clear from this figure that the charge is strongly localized on the Ni atom (center of the figure), with no contribution from the ligands (oxygen). This is a clear case of a positive charge transfer compound, i.e., charge is transferred from the metallic atom towards the ligand. The difference with respect to ligand hole systems is apparent. The existence of ligand holes can be verified in these compounds by looking at the figures shown in Fig 8 b-g: we observe a strong signal on the ligand atoms, indicating a negative charge transfer compound, or the presence of ligand holes. In some cases, mainly for the compounds containing transition metal atoms, there is still a metal-atom component, but the picture is very different from NiO, where absolutely no contribution was observed on the ligands. This is true across different bonding patterns for both *s-p* electron and *d*-electron disproportionated ABX$_3$ compounds. In fact, there is a LH state even in the SLE cases as well, as can be clearly observed by their projected density of states.

***Energy level model for Ligand hole formation***: Figure 7 gives the essential features of *ligand hole formation*. LH formation requires that the relevant metal states should be deeper than ligand orbitals (so electrons can be transferred to the metal) and that a sufficiently strong metal-to-ligand coupling exists so as to create unoccupied hybrid levels with a large fraction of ligand character. Thus, LH does not require disproportionation, but in disproportionated states there exist a short enough B - ligand bond that increases hybridization and consequently increases the ligand character on empty states. These empty levels will have strong oxygen-*p* (or Fluorine or Chlorine) character, showing that the holes are localized on the ligands. This character of the CBM is different from most semiconductor compounds, and is clearly increased by short cation-ligand bonds.

***The driving force for LH formation*** is the *self regulating response*:[39] Total energy lowering favor the formation of LH when without such a LH, the charge on the metal would be highly positive (such as $Ni^{3+}$ in $RNiO_3$ or $Bi^{4+}$ in $BaBiO_3$). This is not favored energetically, so the ligand transfers electrons (thus, forming a hole) to the metal cation so as to self-regulate its charge, creating the [$Ni^{2+}$ - $O^{1-}$] complex in $RNiO_3$ and the [$Bi^{3+}$-$O^{1-}$] complex in $BaBiO_3$ where the hole is on the oxygen octahedra.

.

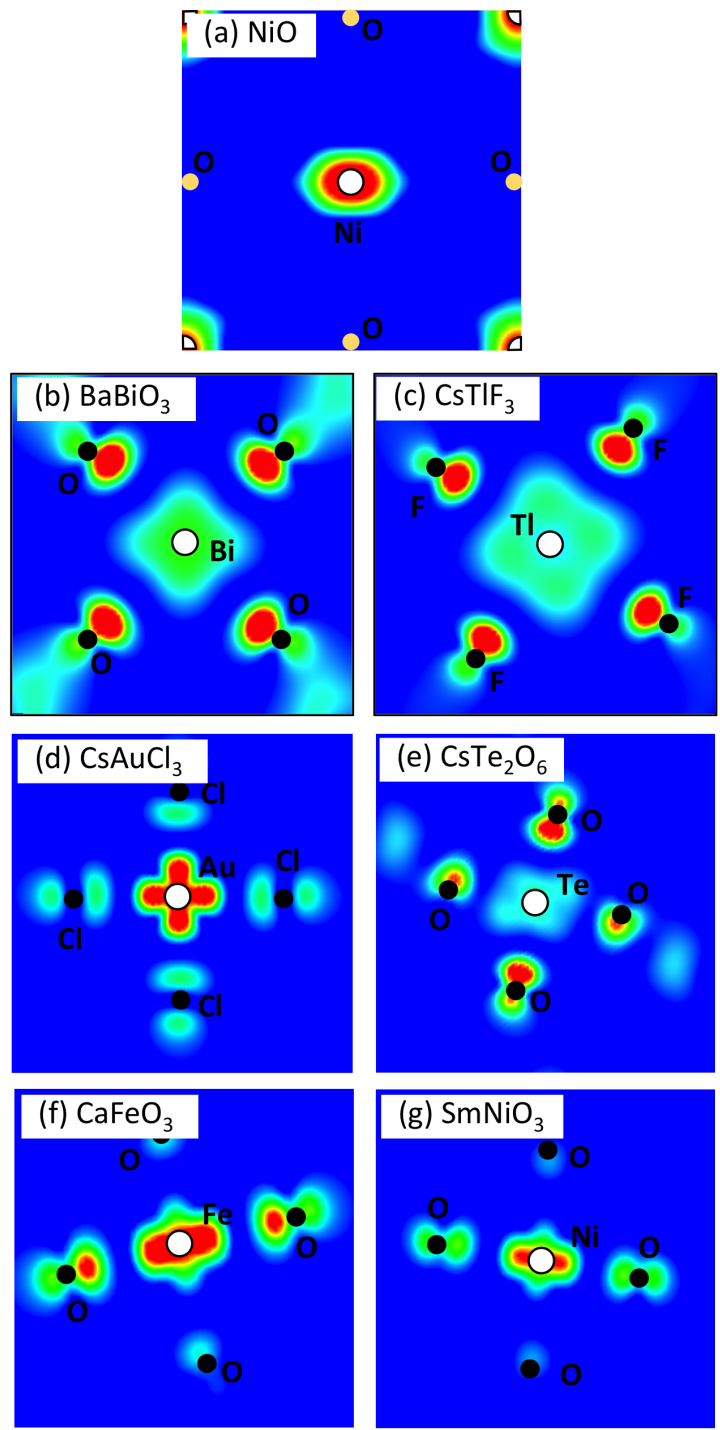

**Figure 8:** Square of the wave function of the ligand hole levels (lowest unoccupied states) for (a) NiO, (b) BaBiO$_3$, (c) CsTlF$_3$, (d) CsAuCl$_3$, (e) CsTe$_2$O$_6$, (f) CaFeO$_3$ and (g) SmNiO$_3$ in a plane containing four ligands and the central B atom in the small octahedra of the DLE phase. The white circles indicate the position of the B atom, and the black/yellow dots indicate the position of the ligands.

## IV. Conclusions

Quantum materials such as transition metal oxide perovskites present a wide range of interesting properties, such as metal insulator transitions, high temperature superconductivity, a variety of magnetic orders, and can exhibit different local environments (DLE) for the same atoms manifested by bond disproportionation. The basic phenomenology associated with such disproportionation involves the absence of B-atom charge ordering, the creation of different B-X bond length ('*bond alternation*') for different local environments, the appearance of metal (in SLE) to insulator (in DLE) transition, and the formation of ligand holes. We point out that:

(i) The broad phenomenology associated with disproportionation is common to a range of chemical bonding patterns in $ABX_3$ compounds, either with s-p electron B-metal cations ($BaBiO_3$, $CsTlF_3$), or noble metal cation ($CsAuCl_3$), as well as d-electron cations ($SmNiO_3$, $CaFeO_3$);

(ii) Using as a guide the lowering of the total energy in DLE relative to SLE, we show that density functional calculations describe this phenomenology across the chemical bonding range without resort to special correlation effects. In particular, lower (DLE) total energy configurations naturally develop bond alternation, gaping of the metallic SLE state, and absence of charge ordering with ligand hole formation;

(iii) Underlying much of this phenomenology is the 'self-regulating response' (SRR), whereby in strongly bonded metal-ligand systems with high lying ligand orbitals, the system protects itself from creating highly charged cations by transferring ligand electrons to the metal, thus preserving a nearly constant metal charge in different local environments, while creating B-ligand bond alternation and ligand-like conduction band ('ligand hole' states).

We address the five questions posed in the Introduction as follows:

(a) DFT provides an adequate level of theory of interelectronic interactions for predicting the tendency of actual $ABX_3$ to be SLE or DLE.

(b) The formal oxidation state is not a physically meaningful 'local environment descriptor' for predicting within the convex hull construct of Fig.1 the tendency of actual $ABX_3$ to be SLE or DLE.

(c) Bond geometry is a physically meaningful descriptor for predicting within the convex hull construct of Fig.1 the tendency of actual $ABX_3$ to be SLE or DLE.

(d) The SLE *vs* DLE selection is directly related to metallic *vs* insulating character of the compound.

(e) Disproportionation per se is not related to ligand hole formation which is a more general phenomena associated with strong metal-ligand bonding for the orbital order of ligand orbital energy being above metal orbital energies. However, in creating a compressed octahedron with short bond lengths, disproportionation provides a platform for ligand hole formation.

**Acknowledgment**

The work at the University of Colorado Boulder was supported by the U.S. Department of Energy, Office of Science, Basic Energy Sciences, Materials Sciences and Engineering Division under Grant No. DE-SC0010467 to C.U.. G.M.D. also thanks financial support from Brazilian agencies FAPESP and CNPq. This work used resources of the National Energy Research Scientific Computing Center, which is supported by the Office of Science of the U.S. Department of Energy under Contract No. DE-AC02-05CH11231. The work at CNRS Thales was supported by the European Research Council grant MINT (contract #615759). Calculations took advantage of the OCCIGEN machines in France through the DARI project EPOC #A0020910084 and of the DECI resources FIONN in Ireland through the PRACE project FiPSCO.